\documentclass[a4paper, 12pt]{article}
\usepackage{mlmodern}
\usepackage{microtype}
\usepackage{amsmath, amssymb}
\usepackage{bm}
\usepackage[dvipsnames]{xcolor}
\usepackage{graphicx}
  \graphicspath{{Figures/}}
\usepackage{subcaption}
\usepackage{enumitem}
\usepackage{titling}
\usepackage{authblk}

\usepackage[
  backend=biber,
  style=phys,
  biblabel=brackets,
  pageranges=false
]{biblatex}
\DeclareFieldFormat[article]{title}{#1}
\DeclareFieldFormat[inproceedings]{title}{#1}
\DeclareFieldFormat[inproceedings]{booktitle}{\mkbibemph{#1}}
\DeclareFieldFormat[book]{title}{\mkbibemph{\href{https://doi.org/\thefield{doi}}{#1}}}
\usepackage[
	hidelinks,
    colorlinks,
    citecolor = Green,
    linkcolor = blue,
    urlcolor  = blue
]{hyperref}

\usepackage[
    left   = 2cm,
    right  = 2cm,
    top    = 2.5cm,
    bottom = 2cm
]{geometry}

\begin{document}

% ------------------------------------------------
\title{Ground states of the Ising model at fixed magnetization on a triangular ladder with three-spin interactions}

\author[,1,2]{Shota Garuchava\thanks{E-mail: shota\_garuchava@hotmail.com}}

\affil[1]{\small\textit{Ilia State University, 0162 Tbilisi, Georgia}}

\affil[2]{\small\textit{Andronikashvili Institute of Physics, 0177 Tbilisi, Georgia}}

\date{\vspace{-2\baselineskip}}

\maketitle

% ------------------------------------------------
\renewcommand{\abstractname}{\vspace{-\baselineskip}}
\begin{abstract}

We study the Ising model at fixed magnetization on a triangular ladder with three-spin interactions.
By recasting the ground-state determination as a linear programming (LP) problem, we solve it exactly using standard LP techniques.
We construct the phase diagram for arbitrary fixed magnetization and identify three types of ground states: periodic, phase-separated, and ordered but aperiodic.
When magnetization is treated as a free parameter, the ground state adopts only periodic configurations with the average magnetization per site $0$, $\pm 1/3$ or $\pm 1$, except for the phase boundaries. 

\end{abstract}
% ------------------------------------------------

% ------------------------------------------------
\section{Introduction}

In the past few decades, experimental techniques to simulate quantum systems using ultracold atoms in optical lattices has been greatly developed~\cite{Bloch_2008, Lewenstein_2007, Bloch_2012, Gross_2017, Esslinger_2010}. 
It enables the realization of vast number of structures with controllable parameters, beyond what is observed in real materials.
The Hubbard model~\cite{Hubbard_1963}, a prototype model in condensed matter physics, can be realized in various geometries~\cite{Becker_2010, Phong_2019, Jo_2012, Soltan_2011, Taruell_2012, Uehlinger_2013, Anisimovas_2016} and with extensions, such as spin-dependent hopping~\cite{Zoller_2004, Taglieber_2008, Jotzu_2015} and synthetic gauge field~\cite{Dalibard_2011, Miyake_2013, Celi_2014, Galitski_2019, Aidelsburger_2013}.

\begin{figure}[t]
    \centering
    \includegraphics[scale=1]{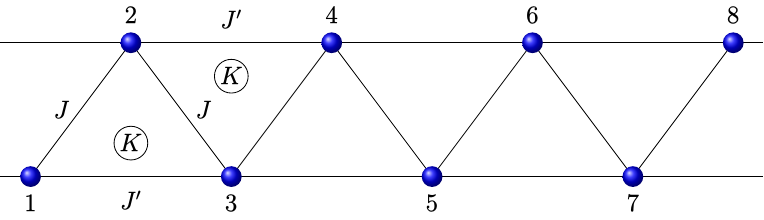}
    \caption{Sketch of the lattice for the model~\eqref{eq:H}. 
    The sites are enumerated along the zig-zag chain. 
    Periodic boundary conditions are imposed, and each leg contains $L$ sites, giving $2L$ sites in total.}
    \label{fig:lattice}
\end{figure}

Recently, an effective spin Hamiltonian describing the spin-asymmetric Hubbard model on a triangular ladder has been derived in the strong-coupling regime~\cite{Garuchva_2025}.
In the Falicov--Kimball limit~\cite{Falicov_Kimball_1969}, where one spin component is localized, it was shown that the effective Hamiltonian reduces to the Ising model with three-spin interactions.
In this work, we study the ground-state phase diagram of the corresponding model, described by the Hamiltonian
% ------------------------------------------------
\begin{align}
\label{eq:H}
    H
    =
    J
    \sum_{i=1}^{2L}
        \sigma_{i}
        \sigma_{i+1}
    +
    J'
    \sum_{i=1}^{2L}
        \sigma_{i}
        \sigma_{i+2}
    +
    K
    \sum_{i=1}^{2L}
        \sigma_{i} \sigma_{i+1} \sigma_{i+2}
\,.
\end{align}
% ------------------------------------------------
Here ${\sigma_i = \pm 1}$. 
The lattice is shown in Fig.~\ref{fig:lattice}. 
Periodic boundary conditions are imposed, and each leg contains $L$ sites, giving $2L$ sites in total. 
In the strong-coupling expansion of the underlying Falicov--Kimball model (FKM), the three-spin interaction emerges at third order, implying that the corresponding coupling is parametrically small. In the present work, however, we go beyond this perturbative regime and investigate the model defined by Eq.~\eqref{eq:H} for arbitrary values of $K$.

The study of generalized Ising models with multi-spin interactions, especially on a frustrated geometries, is a fascinating field of research in its own right.
There are at least two major categories of problems. One is related to thermodynamics~\cite{Ashkin_Teller_1943, Baxter_1972, Baxter_Wu_1973}, and the other to ground state determination, relevant for example in alloy context~\cite{Ceder_1993, van_de_Walle_2002, Sanchez_2010, Widom_2018, Ekborg-Tanner_2024}.
This paper is dedicated to identifying the ground-state configurations at zero temperature at fixed average magnetization per site ${m = (1/2L) \sum_{i=1}^{2L} \sigma_i}$.
Such a formulation of the problem is relevant in the context of ultracold atoms in optical lattices, where the total magnetization of the effective model is determined by the number of particles in the system, which is a fixed quantity.

Numerous exact methods have been developed to determine the ground states of generalized Ising models.
Among them is the method of irreducible blocks for 1D systems with finite-range interactions~\cite{Morita_1974_1, Morita_1974_2, Martinez_2015}. This approach has primarily been applied to unconstrained problems~\cite{Kaburagi_2004}.
The method of ``basic rays" has been employed to solve several 2D~\cite{Dublenych_2009, Dublenych_2013} and 3D~\cite{Dublenych_2022} problems, at least partially. It has also been used to address constrained problems~\cite{Dublenych_2011}, albeit in an indirect manner.
Another technique combines maximum satisfiability and convex optimization~\cite{MAX-SAT_2016}, which allows efficient and provable determination of ground states.
Finally, the linear programming (LP) approach~\cite{Kaburagi_1975, Ducastelle_1992} provides a robust and systematic framework for identifying the complete set of ground states. Its primary shortcoming lies in computational scalability, as it becomes intractable for complex 3D systems. Besides, it is known to yield ``inconstructible" vertices, i.e.\ solutions which do not correspond to realizable configurations~\cite{Ceder_1994}.
In this work, we employ the LP approach to determine the complete phase diagram, that includes resolving the issue of ``inconstructible" vertices for the model~\eqref{eq:H}.

Controlled multi-spin Ising-type interactions have been experimentally engineered in trapped-ion quantum simulators, see Ref.~\cite{Katz_2023} and references therein. The resulting classical spin configurations are detected via state-dependent fluorescence of individual ions~\cite{Todaro_2021, Bruzewicz_2019}.

The paper is organized as follows. 
In Sec.~\ref{sec:energy}, average energy per site is discussed, together with the way states are classified. 
In Sec.~\ref{sec:constraints}, structural constraints are provided, including the ones due to finite lattice. 
In Sec.~\ref{sec:parametrization}, the parametrization is introduced, while its derivation is presented in App.~\ref{app:parametrization}. These parameters constitute the decision variables of the LP, described in subsequent Sec.~\ref{sec:LP}.
The phase diagram is discussed in Sec.~\ref{sec:PD} and the summary is provided in Sec.~\ref{sec:summary}.

% ------------------------------------------------
% ------------------------------------------------

% ------------------------------------------------
\section{Energy and state enumeration}
\label{sec:energy}

From Eq.~\eqref{eq:H}, it is evident that the average energy per site has the form
% ------------------------------------------------
\begin{gather}
\label{eq:energy}
    \varepsilon (\bm{x} ; \bm{c})
    :=
    \frac{E (\bm{x} ; \bm{c})}{2L}
    =
    \bm{c} \cdot \bm{x}
\,,
\qquad
    \bm{c} = \begin{pmatrix}J & J' & K \end{pmatrix}
,
\end{gather}
% ------------------------------------------------
where $E (\bm{x} ; \bm{c})$ is the total energy in a given state, $\bm{c}$ is the parameter vector and $\bm{x}$ is the coefficient vector, where $x_i$ with ${-1 \leqslant x_i \leqslant 1}$ denotes the rational coefficient of the parameter $c_i$.
The vector $\bm{x}$ defines a state, or in general a set of degenerate states.
Our goal is to minimize ${\varepsilon (\bm{x} ; \bm{c})}$ with respect to $\bm{x}$ for any given $\bm{c}$. The problem lies in the fact that the components of $\bm{x}$ are not all independent. Their values must correspond to a physically realizable spin configuration.

\begin{figure}[t!]
    \centering
    \includegraphics[scale=1]{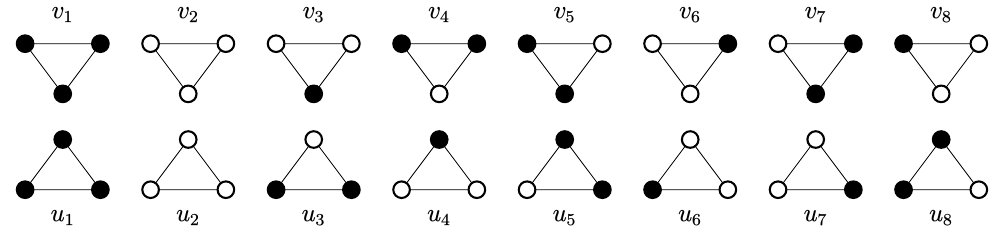}
        \caption{All possible spin configurations on triangular plaquettes. Solid and open circles represent ${\sigma = +1}$ and ${\sigma = -1}$, respectively. The symbols $u_i$ and $v_i$ label the configurations and also denote their occurrence frequencies ${(0 \leqslant u_i, v_i \leqslant 1)}$ in a given state.
        }
    \label{fig:triangles}
\end{figure}

The lattice consists of two types of triangular plaquettes, which we refer to as $u$- and $v$-triangles. A $u$-triangle has two vertices on the lower leg and one on the upper, whereas a $v$-triangle has two vertices on the upper leg and one on the lower. All possible spin configurations are listed in Fig.~\ref{fig:triangles}.
We classify states by a set of 16 normalized numbers
% ------------------------------------------------
\begin{align}
    u_i = \frac{N^{(u)}_i}{L} \,,\quad
    v_i = \frac{N^{(v)}_i}{L} \,,\quad
    (i = 1,2,\ldots,8) \,,
\end{align}
% ------------------------------------------------
where $N^{(u)}_i$ and $N^{(v)}_i$ denote the total numbers of $u$- and $v$-triangles in the $i^{\mathrm{th}}$ configuration of a given state, respectively.
We shall slightly abuse notation by using $u_i$ and $v_i$ to denote both the configurations and their corresponding frequencies, the intended meaning being clear from the context.
It is straightforward to express the components of $\bm{x}$ in terms of $u_i$ and $v_i$
% ------------------------------------------------
\begin{align}
\label{eq:x-uv}
\begin{alignedat}{4}
    &(\text{Coefficient of } J&&){:}\quad
    &&x_1
    &&=
    \left(u_1 + u_2\right)
    -
    \left(u_3 + u_4\right)
,
\\
    &(\text{Coefficient of } J'&&){:}\quad
    &&x_2
    &&=
    \frac{1}{2}
    \left(u_1 + u_2 + u_3 + u_4\right)
    -
    \frac{1}{2}
    \left(u_5 + u_6 + u_7 + u_8\right)
\\
    & && && &&
    \qquad
    +
    \frac{1}{2}
    \left(v_1 + v_2 + v_3 + v_4\right)
    -
    \frac{1}{2}
    \left(v_5 + v_6 + v_7 + v_8\right)
,
\\
    &(\text{Coefficient of } K&&){:}\quad
    &&x_3
    &&=
    \frac{1}{2}
    \left(u_1 + u_4 + u_6 + u_7\right)
    -
    \frac{1}{2}
    \left(u_2 + u_3 + u_5 + u_8\right)
\\
    & && && &&
    \qquad
    +
    \frac{1}{2}
    \left(v_1 + v_3 + v_6 + v_8\right)
    -
    \frac{1}{2}
    \left(v_2 + v_4 + v_5 + v_7\right)
.
\end{alignedat}
\end{align}
% ------------------------------------------------
For illustration, we derive $x_1$ explicitly. 
The $u_1$ and $u_2$ configurations each contribute $+2J$ to the total energy, while $u_3$ and $u_4$ each contribute $-2J$.
Since the total number of $u$-triangles is $L$, which is half the number of sites, this corresponds to an energy of $\pm J$ per site. 
The configurations $u_5$, $u_6$, $u_7$ and $u_8$ do not contribute $J$ to the energy.
Equivalently, $x_1$ could be computed by counting only $v$-triangles.
Remaining components of $\bm{x}$ can be obtained by the same reasoning.

% ------------------------------------------------
% ------------------------------------------------

% ------------------------------------------------
\section{Constraints}
\label{sec:constraints}

The constraints among $x_i$ are not immediately apparent. However, they can be identified for $u_i$ and $v_i$, and then translated into constraints on $x_i$ by properly inverting Eq.~\eqref{eq:x-uv}.
The constraints in terms of $u_i$ and $v_i$ are as follows.
First, the total number of triangles of each type must equal $L$, hence the frequencies
% ------------------------------------------------
\begin{align}
\label{eq:sum_uv}
    \sum_{i=1}^8 u_i
    =
    \sum_{i=1}^8 v_i
    =
    1
\,.
\end{align}
% ------------------------------------------------
Next, two neighboring $u$- and $v$-triangles have a shared edge. Therefore, the number of $u$-triangles with a given spin configuration on the right edge must equal the number of $v$-triangles with the same spin configuration on the left edge
% ------------------------------------------------
\begin{align}
\begin{aligned}
\label{eq:bot_R-top-L}
    u_1 + u_5 = v_1 + v_5 \,,\qquad
    u_2 + u_6 = v_2 + v_6 \,,\\
    u_3 + u_7 = v_3 + v_7 \,,\qquad
    u_4 + u_8 = v_4 + v_8 \,.
\end{aligned}
\end{align}
% ------------------------------------------------
Conversely, the number of $u$-triangles with a given spin configuration on the left edge must equal the number of $v$-triangles with the same spin configuration on the right edge
% ------------------------------------------------
\begin{align}
\begin{aligned}
\label{eq:bot_L-top-R}
    u_1 + u_8 = v_1 + v_7 \,,\qquad
    u_2 + u_7 = v_2 + v_8 \,,\\
    u_3 + u_6 = v_3 + v_5 \,,\qquad
    u_4 + u_5 = v_4 + v_6 \,.
\end{aligned}
\end{align}
% ------------------------------------------------
Eight of the 10 relations in Eqs.~\eqref{eq:sum_uv}, \eqref{eq:bot_R-top-L} and \eqref{eq:bot_L-top-R} are linearly independent.
However, the selection of these 8 is not unique.
This leaves 8 out of the 16 degrees of freedom.
Fixing the magnetization provides one additional independent constraint. There are several ways to express it in terms of $u_i$ and $v_i$, one possible form is
% ------------------------------------------------
\begin{align}
\label{eq:m}
    m
    &=
    (u_1 - u_2) + \frac{1}{2} (u_5 + u_8) - \frac{1}{2} (u_6 + u_7)
\,.
\end{align}
% ------------------------------------------------
Fixing the magnetization freezes one of the 8 degrees of freedom.

Finally, all of these constraints can be satisfied with negative $u_i$ and $v_i$, which is not physical. Therefore, we require
% ------------------------------------------------
\begin{align}
\label{eq:non-negative-uv}
    u_i, v_i \geqslant 0 \,.
\end{align}
% ------------------------------------------------
The normalization condition \eqref{eq:sum_uv} together with non-negativity automatically ensures the upper bounds ${u_i, v_i \leqslant 1}$.

% ------------------------------------------------
\subsection{Finite-size constraints}
\label{sec:finite-size}

The conditions in Eqs.~\eqref{eq:sum_uv}, \eqref{eq:bot_R-top-L} and \eqref{eq:bot_L-top-R} are necessary but not sufficient for physical realizability. 
For a finite lattice, additional logical constraints must be imposed.
For example, if ${u_1 + u_3}$ and ${u_2 + u_4}$ are both nonzero, then one must also require ${u_5 + u_7 \geqslant 1/L}$ and ${u_6 + u_8 \geqslant 1/L}$, which are not implied by Eqs.~\eqref{eq:sum_uv}, \eqref{eq:bot_R-top-L} and \eqref{eq:bot_L-top-R}.
In the thermodynamic limit, where ${1/L \to 0}$, all such constraints can be neglected.
However, they leave an imprint, which will be discussed in Sec.~\ref{sec:PD}.

% ------------------------------------------------
\section{Parametrization}
\label{sec:parametrization}

We aim to invert Eq.~\eqref{eq:x-uv} in a manner that satisfies all the constraints given by Eqs.~\eqref{eq:sum_uv}, \eqref{eq:bot_R-top-L}, \eqref{eq:bot_L-top-R} and \eqref{eq:m}.
The number of components of the vector $\bm{x}$ is ${d_x = 3}$, while $u_i$ and $v_i$ combined ${(i=1,2,\ldots,8)}$ have 7 degrees of freedom, as was shown in Sec.~\ref{sec:constraints}. Therefore, in order to invert Eq.~\eqref{eq:x-uv} while preserving all degrees of freedom, we must introduce the auxiliary vector $\bm{y}$ with ${d_y = 7 - d_x = 4}$ components. 
Although the complete auxiliary vector carries unique physical content, its individual components may admit slightly different interpretations depending on the chosen basis. We choose the basis such that
$\bm{y}$ quantifies the following:
$y_1$ -- difference of $J$ contributions between the positive- and negative-slope rungs (the left and right edges of the $u$-triangles, respectively);
$y_2$ -- difference of $J'$ contributions between the lower and upper legs;
$y_3$ -- difference of $K$ contributions between the $u$- and $v$-triangles;
and $y_4$ -- difference of the magnetization between the lower and upper legs.
In that case, Eq.~\eqref{eq:x-uv} is inverted as
% ------------------------------------------------
{\small
\begin{align}
\begin{alignedat}{3}
\label{eq:parametrization}
    u_1 &= \frac{1 + 3m + 2x_1 + x_2 + y_2 + x_3 + y_3 + \hphantom{3}y_4 }{8} , \quad
    &&v_1 &&= \frac{1 + 3m + 2x_1 + x_2 - y_2 + x_3 - y_3 - \hphantom{3}y_4}{8} , \\
    u_2 &= \frac{1 - 3m + 2x_1 + x_2 + y_2 - x_3 - y_3 - \hphantom{3}y_4}{8} , \quad
    &&v_2 &&= \frac{1 - 3m + 2x_1 + x_2 - y_2 - x_3 + y_3 + \hphantom{3}y_4}{8} , \\
    u_3 &= \frac{1 + \hphantom{3}m - 2x_1 + x_2 + y_2 - x_3 - y_3 + 3y_4}{8} , \quad
    &&v_3 &&= \frac{1 - \hphantom{3}m - 2x_1 + x_2 - y_2 + x_3 - y_3 + 3y_4}{8} , \\
    u_4 &= \frac{1 - \hphantom{3}m - 2x_1 + x_2 + y_2 + x_3 + y_3 - 3y_4}{8} , \quad
    &&v_4 &&= \frac{1 + \hphantom{3}m - 2x_1 + x_2 - y_2 - x_3 + y_3 - 3y_4}{8} , \\
    u_5 &= \frac{1 + \hphantom{3}m - 2y_1 - x_2 - y_2 - x_3 - y_3 - \hphantom{3}y_4}{8} , \quad
    &&v_5 &&= \frac{1 + \hphantom{3}m - 2y_1 - x_2 + y_2 - x_3 + y_3 + \hphantom{3}y_4}{8} , \\
    u_6 &= \frac{1 - \hphantom{3}m - 2y_1 - x_2 - y_2 + x_3 + y_3 + \hphantom{3}y_4}{8} , \quad
    &&v_6 &&= \frac{1 - \hphantom{3}m - 2y_1 - x_2 + y_2 + x_3 - y_3 - \hphantom{3}y_4}{8} , \\
    u_7 &= \frac{1 - \hphantom{3}m + 2y_1 - x_2 - y_2 + x_3 + y_3 + \hphantom{3}y_4}{8} , \quad
    &&v_7 &&= \frac{1 + \hphantom{3}m + 2y_1 - x_2 + y_2 - x_3 + y_3 + \hphantom{3}y_4}{8} , \\
    u_8 &= \frac{1 + \hphantom{3}m + 2y_1 - x_2 - y_2 - x_3 - y_3 - \hphantom{3}y_4}{8} , \quad
    &&v_8 &&= \frac{1 - \hphantom{3}m + 2y_1 - x_2 + y_2 + x_3 - y_3 - \hphantom{3}y_4}{8} .
\end{alignedat}
\end{align}
}%
% ------------------------------------------------
See the derivation in App.~\ref{app:parametrization}.
The vector $\bm{y}$ is a variable that does not appear in the objective function $\varepsilon(\bm{x};c)$. However, it is essential for a complete classification of the states and consequently for the minimization problem.

% ------------------------------------------------
\section{Linear programming}
\label{sec:LP}

To simplify the presentation, we define a vector $\bm{w}$ with ${d_w=16}$ components
% ------------------------------------------------
\begin{align}
\label{eq:w}
    \bm{w} := 
    \begin{pmatrix}
    \bm{u} & \bm{v}
    \end{pmatrix}^{\mathrm{T}}
,
\end{align}
% ------------------------------------------------
which combines all entries of ${\bm{u} := (u_1,u_2,\ldots,u_8)}$ and ${\bm{v} := (v_1,v_2,\ldots,v_8)}$ into a single symbol.
This notation allows us to treat $\bm{u}$ and $\bm{v}$ simultaneously on an equal footing.

The equality constraints given by Eqs.~\eqref{eq:sum_uv}, \eqref{eq:bot_R-top-L}, \eqref{eq:bot_L-top-R} and \eqref{eq:m} are automatically satisfied by the parametrization~\eqref{eq:parametrization}.
The only remaining constraints are inequalities, specifically ${w_i \geqslant 0}$, which translate into linear inequalities in $\bm{x}$ and $\bm{y}$.
Minimization of the linear function $\varepsilon(\bm{x}; \bm{c})$ subject to such linear inequality constraints is known as the \emph{linear programming} (LP).
In our case, $\bm{y}$ is treated as a decision variable with zero weight in the objective function.

We apply standard LP tools and techniques to solve the problem.
Taking the thermodynamic limit significantly reduces the computational effort. 
For a finite lattice, additional logical constraints must be imposed, as noted in Sec.~\ref{sec:finite-size}.
The inclusion of such logical conditions effectively turns the LP into a mixed-integer LP, which belongs to a higher computational complexity class.
It is easier to neglect these constraints and analyze their residual effect in the result.

The ${d_w=16}$ inequality constraints ${w_i(\bm{x},\bm{y}) \geqslant 0}$ define the feasible region, which is a convex polytope in the ${(d_x+d_y)}$-dimensional variable space (7D).
The global minimum is attained at one of its vertices, or on a face in case of degeneracy.
Each vertex corresponds to an $m$-dependent point in the variable space, which satisfy at least ${d_x+d_y=7}$ inequality constraints as equalities.
At every vertex, the components of $\bm{w}^\mu(m)$ has the form ${w^\mu_i(m) = \alpha^\mu_i + \beta^\mu_i m}$, where $\mu$ enumerates the vertices, and $\alpha^\mu_i, \beta^\mu_i$ are rational numbers.
Because ${0 \leqslant w^\mu_i(m) \leqslant 1}$, each vertex solution is admissible only within a certain interval ${m_1 \leqslant m \leqslant m_2}$.
Consequently, the range of $m$ is partitioned into intervals, each corresponding to a fixed set of admissible vertices. The phase diagram is then constructed separately for those intervals.
For the model~\eqref{eq:H}, we find that the range ${-1 \leqslant m \leqslant 1}$ is divided into 4 intervals by the critical values ${m = 0}$, ${m = \pm 1/3}$ and ${m = \pm 1}$.

Not every vertex corresponds to the ground state for some parameter vector $\bm{c}$. This would hold only if every component of $\bm{y}$ entered the objective function with nonzero weight.
Since the Hamiltonian~\eqref{eq:H} lacks the corresponding terms, one must filter the vertices. This is done by projecting the vertices of the ${(d_x+d_y)}$-dimensional feasible polytope (7D) onto the $d_x$-dimensional subspace (3D) spanned by $\bm{x}$, which simply amounts to discarding the $\bm{y}$-components of the vertex coordinates. 
Then, the convex hull of the projected points is constructed. The vertices of this hull correspond to the ground states attainable for some $\bm{c}$.

% ------------------------------------------------
\section{Phase diagram}
\label{sec:PD}

The solution of the LP problem yields a set of $m$-dependent vectors $\bm{w}^\mu(m)$.
Without loss of generality, we restrict our analysis to ${m \geqslant 0}$.
We treat the cases ${J>0}$ and ${J<0}$ separately and present the corresponding phase diagrams in the ${(K,J')}$ plane for fixed $J$.

We first construct the phase diagrams for the critical magnetization values ${m=0}$ and ${m=1/3}$.
For these cases, the components of the $\bm{w}^\mu$ vectors become rational numbers.
For each $\bm{w}^\mu$, the corresponding energy can be directly evaluated using Eqs.~\eqref{eq:energy} and \eqref{eq:x-uv}, while identifying the state requires additional analysis.
Specifically, one must determine which $u$-$v$ configurations can appear adjacent to each other:
% ------------------------------------------------
\begin{align}
\begin{alignedat}{4}
    {u_1, u_5} &\to {v_1, v_5}\,,\quad
    {u_2, u_6} &&\to {v_2, v_6}\,,\quad
    {u_3, u_7} &&\to {v_3, v_7}\,,\quad
    {u_4, u_8} &&\to {v_4, v_8}\,,
\\
    {v_1, v_7} &\to {u_1, u_8}\,,\quad
    {v_2, v_8} &&\to {u_2, u_7}\,,\quad
    {v_3, v_5} &&\to {u_3, u_6}\,,\quad
    {v_4, v_6} &&\to {u_4, u_5}\,,
\end{alignedat}
\end{align}
% ------------------------------------------------
indicating, for example, that $u_1$ and $u_5$ can be followed on the right by either $v_1$ or $v_5$, and so forth (see Fig.~\ref{fig:triangles}).
Each $\bm{w}^\mu$ might describe one of the three scenarios: 1) a single cyclic $u$-$v$ sequence, corresponding to a periodic ground state---generated by a single repeating finite supercell; 2) two disjoint cyclic $u$-$v$ sequences, corresponding to a phase-separated ground state---splitting the system into two parts, each forming a periodic structure generated by a distinct repeating supercell; and 3) multiple admissible $u$-$v$ sequences, corresponding to an ordered but aperiodic ground state---consisting of specific blocks in specified amounts, but arranged arbitrarily subject to certain local rules.

Each state with a period of $2r$ sites is $r$-fold degenerate due to translational symmetry, while phase-separated and aperiodic states exhibit linear and combinatorial degeneracy with system size, respectively.

For ${m=0}$, there are only periodic and phase-separated ground states, while ${m=1/3}$ exhibits all three kinds of ground states.
Below we present the complete solution to the LP problem.

% ------------------------------------------------
\subsubsection*{Periodic ground states ${(m=0)}$}

% ------------------------------------------------
\begin{enumerate}

% ------------------------------------------------
\item 
$\left(u_3 = 1; \ v_3 = 1\right)$
\, or \,
$\left(u_4 = 1; \ v_4 = 1\right)$:
% ------------------------------------------------
\begin{align}
    \varepsilon = - J + J^{\prime}
\,,\qquad
    \psi
    =
    \left(
        \raisebox{-0.75ex}{$\bullet$}
        \raisebox{ 0.75ex}{$\circ$}
    \right)^{\otimes L}
\quad
\text{or}
\quad
    \psi
    =
    \left(
        \raisebox{-0.75ex}{$\circ$}
        \raisebox{ 0.75ex}{$\bullet$}
    \right)^{\otimes L}
.
\end{align}
% ------------------------------------------------

% ------------------------------------------------
\item 
$(u_7 = u_8 = 1/2; \ v_7 = v_8 = 1/2)$
\, or \,
$(u_5 = u_6 = 1/2; \ v_5 = v_6 = 1/2)$:
% ------------------------------------------------
\begin{align}
    \varepsilon = - J^{\prime}
\,,\qquad
    \psi
    =
    \left(
        \raisebox{-0.75ex}{$\bullet$}
        \raisebox{ 0.75ex}{$\bullet$}
        \raisebox{-0.75ex}{$\circ$}
        \raisebox{ 0.75ex}{$\circ$}
    \right)^{\otimes L/2}
\quad
\text{or}
\quad
    \psi
    =
    \left(
        \raisebox{ 0.75ex}{$\bullet$}
        \raisebox{-0.75ex}{$\bullet$}
        \raisebox{ 0.75ex}{$\circ$}
        \raisebox{-0.75ex}{$\circ$}
    \right)^{\otimes L/2}
.
\end{align}
% ------------------------------------------------

\end{enumerate}
% ------------------------------------------------

% ------------------------------------------------
\subsubsection*{Phase-separated ground states ${(m=0)}$}

% ------------------------------------------------
\begin{enumerate}

% ------------------------------------------------
\item 
$(u_1 = u_2 = 1/2; \ v_1 = v_2 = 1/2)$:
% ------------------------------------------------
\begin{align}
\label{eq:PS-ex}
    \varepsilon
    =
    J + J^{\prime}
\,,\qquad
    \psi
    =
    \left(
        \raisebox{-0.75ex}{$\bullet$}
        \raisebox{ 0.75ex}{$\bullet$}
    \right)^{\otimes L/2}
    \otimes
    \left(
        \raisebox{-0.75ex}{$\circ$}
        \raisebox{ 0.75ex}{$\circ$}
    \right)^{\otimes L/2}
.
\end{align}
% ------------------------------------------------

% ------------------------------------------------
\item 
$(u_1 = u_4 = u_6 = u_7 = 1/4; \ 
v_1 = v_3 = v_6 = v_8 = 1/4)$:
% ------------------------------------------------
\begin{subequations}
\begin{align}
    \varepsilon
    =
    \hphantom{-} K
\,,\qquad
    \psi
    =
    \left(
        \raisebox{-0.75ex}{$\bullet$}
        \raisebox{ 0.75ex}{$\bullet$}
    \right)^{\otimes L/4}
    \otimes
    \left(
        \raisebox{-0.75ex}{$\circ$}
        \raisebox{ 0.75ex}{$\bullet$}
        \raisebox{-0.75ex}{$\circ$}
        \raisebox{ 0.75ex}{$\circ$}
        \raisebox{-0.75ex}{$\bullet$}
        \raisebox{ 0.75ex}{$\circ$}
    \right)^{\otimes L/4}
,
% ------------------------------------------------
\intertext{
$(u_2 = u_3 = u_5 = u_8 = 1/4; \
v_2 = v_4 = v_5 = v_7 = 1/4)$:
}
% ------------------------------------------------
    \varepsilon
    =
    - K
\,,\qquad
    \psi
    =
    \left(
        \raisebox{-0.75ex}{$\circ$}
        \raisebox{ 0.75ex}{$\circ$}
    \right)^{\otimes L/4}
    \otimes
    \left(
        \raisebox{-0.75ex}{$\bullet$}
        \raisebox{ 0.75ex}{$\circ$}
        \raisebox{-0.75ex}{$\bullet$}
        \raisebox{ 0.75ex}{$\bullet$}
        \raisebox{-0.75ex}{$\circ$}
        \raisebox{ 0.75ex}{$\bullet$}
    \right)^{\otimes L/4}
.
\end{align}
\end{subequations}
% ------------------------------------------------

\end{enumerate}
% ------------------------------------------------
There is one subtlety concerning the phase-separated ground states. As an example, consider Eq.~\eqref{eq:PS-ex}, where the LP solution gives $u_1 = u_2 = v_1 = v_2 = 1/2$. No microscopic configuration can realize these frequencies exactly. This solution describes two disjoint cyclic chains, one constructed by using ${u_1 = v_1 =1/2}$ only, and the other one by ${u_2 = v_2 =1/2}$.
In the thermodynamic limit, for a chain model, one resolves the issue by manually introducing two domain walls between the two cyclic chains.
As a result, the energy per site changes only infinitesimally.
In fact, for the illustrated state in Eq.~\eqref{eq:PS-ex}, the frequencies are $u_1 = u_2 = v_1 = v_2 = 1/2 - 1/L$ and $u_7 = u_8 = v_7 = v_8 = 1/L$. The $\mathcal{O}(1/L)$ corrections are then neglected both in $u_i, v_i$ and in the corresponding energy per site.
This kind of discrepancy is referred to as ``inconstructible" vertices in the literature. It arises because the logical constraints, discussed in Sec.~\ref{sec:finite-size}, are not explicitly accounted for. 
Only one representative configuration is shown in each case. We do not specify which domain walls have the lowest energy, as they differ infinitesimally and for our purposes it can be neglected.

% ------------------------------------------------
\subsubsection*{Periodic ground states ${(m=1/3)}$}

% ------------------------------------------------
\begin{enumerate}

% ------------------------------------------------
\item 
$(u_3 = u_5 = u_8 = 1/3; \
v_4 = v_5 = v_7 = 1/3)$:
% ------------------------------------------------
\begin{align}
    \varepsilon
    =
    - \frac{J}{3} - \frac{J^{\prime}}{3} - K
\,,\qquad
    \psi
    =
    \left(
        \raisebox{-0.75ex}{$\bullet$}
        \raisebox{ 0.75ex}{$\circ$}
        \raisebox{-0.75ex}{$\bullet$}
        \raisebox{ 0.75ex}{$\bullet$}
        \raisebox{-0.75ex}{$\circ$}
        \raisebox{ 0.75ex}{$\bullet$}
    \right)^{\otimes L/3}
.
\end{align}
% ------------------------------------------------

\end{enumerate}
% ------------------------------------------------

% ------------------------------------------------
\subsubsection*{Phase-separated ground states ${(m=1/3)}$}

% ------------------------------------------------
\begin{enumerate}

% ------------------------------------------------
\item 
$(u_1 = 2/3,\; u_2 = 1/3; \
v_1 = 2/3,\; v_2 = 1/3)$:
% ------------------------------------------------
\begin{align}
    \varepsilon
    =
    J + J^{\prime} + \frac{K}{3}
\,,\qquad
    \psi
    =
    \left(
        \raisebox{-0.75ex}{$\bullet$}
        \raisebox{ 0.75ex}{$\bullet$}
    \right)^{\otimes 2L/3}
    \otimes
    \left(
        \raisebox{-0.75ex}{$\circ$}
        \raisebox{ 0.75ex}{$\circ$}
    \right)^{\otimes L/3}
.
\end{align}
% ------------------------------------------------

% ------------------------------------------------
\item 
$(u_1 = 1/3,\; u_3 = 2/3; \
v_1 = 1/3,\; v_3 = 2/3)$
\, or \,
$(u_1 = 1/3,\; u_4 = 2/3; \
v_1 = 1/3,\; v_4 = 2/3)$:
% ------------------------------------------------
\begin{align}
    \varepsilon
    =
    - \frac{J}{3} + J^{\prime} + \frac{K}{3}
\,,\quad
    \psi
    =
    \left(
        \raisebox{-0.75ex}{$\bullet$}
        \raisebox{ 0.75ex}{$\bullet$}
    \right)^{\otimes L/3}
    \otimes
    \left(
        \raisebox{-0.75ex}{$\bullet$}
        \raisebox{ 0.75ex}{$\circ$}
    \right)^{\otimes 2L/3}
\quad
\text{or}
\quad
    \psi
    =
    \left(
        \raisebox{-0.75ex}{$\bullet$}
        \raisebox{ 0.75ex}{$\bullet$}
    \right)^{\otimes L/3}
    \otimes
    \left(
        \raisebox{-0.75ex}{$\circ$}
        \raisebox{ 0.75ex}{$\bullet$}
    \right)^{\otimes 2L/3}
.
\end{align}
% ------------------------------------------------

% ------------------------------------------------
\item 
$(u_1 = 1/2,\; u_4 = u_6 = u_7 = 1/6; \
v_1 = 1/2,\; v_3 = v_6 = v_8 = 1/6)$:
% ------------------------------------------------
\begin{align}
    \varepsilon
    =
    \frac{J}{3} + \frac{J^{\prime}}{3} + K
\,,\qquad
    \psi
    =
    \left(
        \raisebox{-0.75ex}{$\bullet$}
        \raisebox{ 0.75ex}{$\bullet$}
    \right)^{\otimes L/2}
    \otimes
    \left(
        \raisebox{-0.75ex}{$\circ$}
        \raisebox{ 0.75ex}{$\bullet$}
        \raisebox{-0.75ex}{$\circ$}
        \raisebox{ 0.75ex}{$\circ$}
        \raisebox{-0.75ex}{$\bullet$}
        \raisebox{ 0.75ex}{$\circ$}
    \right)^{\otimes L/6}
.
\end{align}
% ------------------------------------------------

\end{enumerate}
% ------------------------------------------------

% ------------------------------------------------
\subsubsection*{Aperiodic ground states ${(m=1/3)}$}

% ------------------------------------------------
\begin{enumerate}

% ------------------------------------------------
\item 
$(u_1 = u_7 = u_8 = 1/3; \
v_1 = v_7 = v_8 = 1/3)$
\, or \,
$(u_1 = u_5 = u_6 = 1/3; \
v_1 = v_5 = v_6 = 1/3)$:
% ------------------------------------------------
\begin{align}
    \varepsilon
    =
    \frac{J}{3} - \frac{J^{\prime}}{3} + \frac{K}{3}
\,,\qquad
    \psi
    =
    \frac{2L}{3}
    \left(
        \raisebox{-0.75ex}{$\bullet$}
        \raisebox{ 0.75ex}{$\bullet$}
    \right)
    +
    \frac{L}{3}\,
    \overline{
    \left(
        \raisebox{-0.75ex}{$\circ$}
        \raisebox{ 0.75ex}{$\circ$}
    \right)
    }
\quad
\text{or}
\quad
    \psi
    =
    \frac{2L}{3}
    \left(
        \raisebox{ 0.75ex}{$\bullet$}
        \raisebox{-0.75ex}{$\bullet$}
    \right)
    +
    \frac{L}{3}\,
    \overline{
    \left(
        \raisebox{ 0.75ex}{$\circ$}
        \raisebox{-0.75ex}{$\circ$}
    \right)
    }
.
\end{align}
% ------------------------------------------------

\end{enumerate}%
% ------------------------------------------------
Here we adopt the notation $\psi = n_1 (B_1) + n_2 \overline{(B_2)}$ to denote a configuration composed of $n_1$ blocks of type $B_1$ and $n_2$ blocks of type $B_2$. 
The plus sign signifies coexistence of different block types within the configuration, rather than algebraic addition.
The overline indicates that blocks of type $B_2$ are placed such that no two of them are adjacent. 
The blocks can be arranged in multiple inequivalent ways without violating the non-adjacency constraint, giving the same total energy and frequencies $u_i, v_i$. This degeneracy renders the state aperiodic.

% ------------------------------------------------
\begin{figure}[b!]
    \begin{subfigure}[b]{0.45\textwidth}
         \centering
         \includegraphics[width=\textwidth]{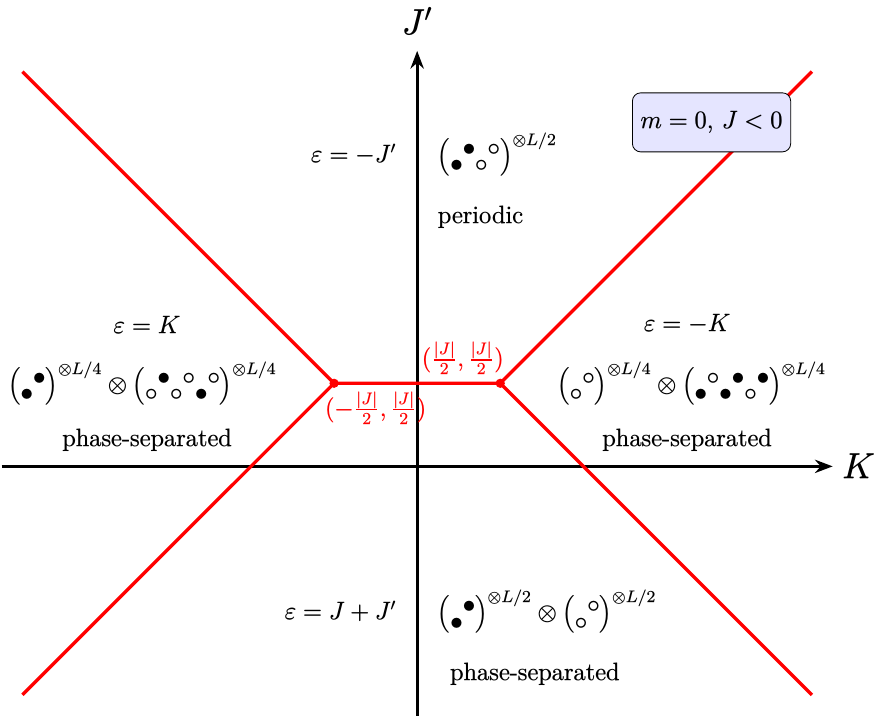}
    \end{subfigure}
    \hfill
    \begin{subfigure}[b]{0.45\textwidth}
         \centering
         \includegraphics[width=\textwidth]{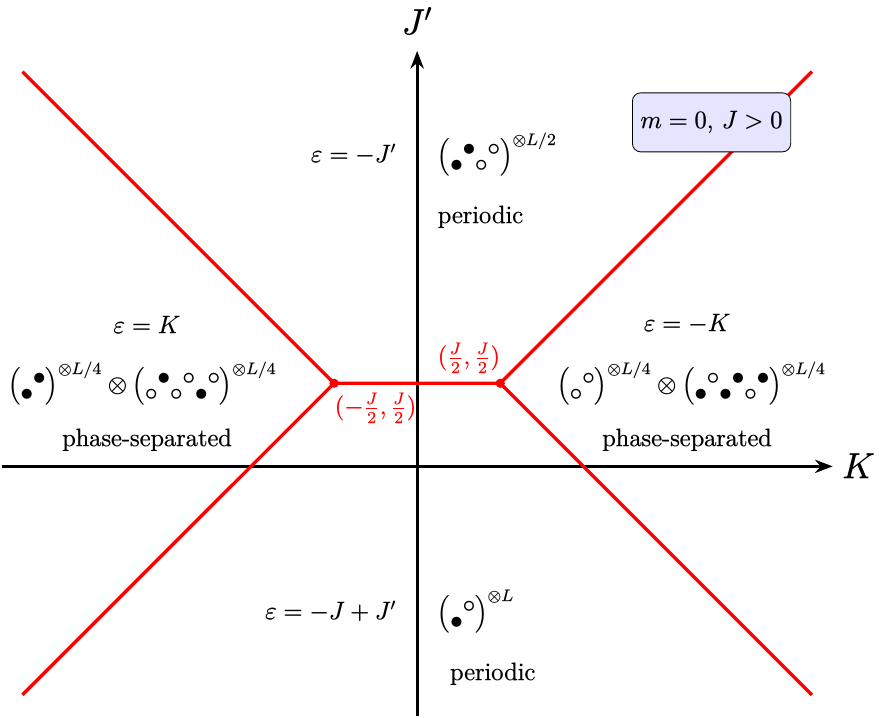}
    \end{subfigure}
\bigbreak
    \begin{subfigure}[b]{0.45\textwidth}
         \centering
         \includegraphics[width=\textwidth]{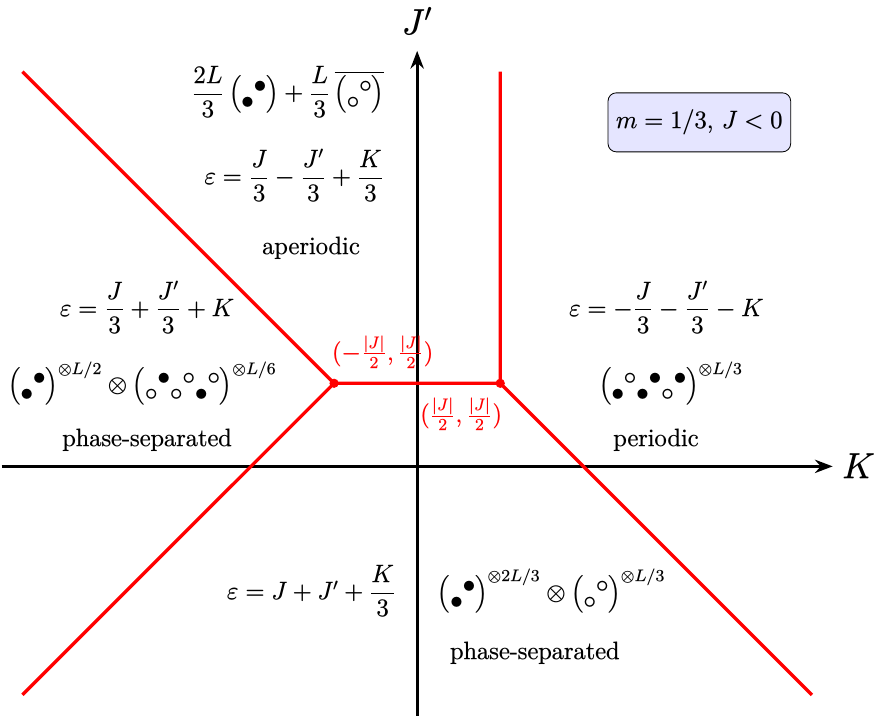}
    \end{subfigure}
    \hfill
    \begin{subfigure}[b]{0.45\textwidth}
         \centering
         \includegraphics[width=\textwidth]{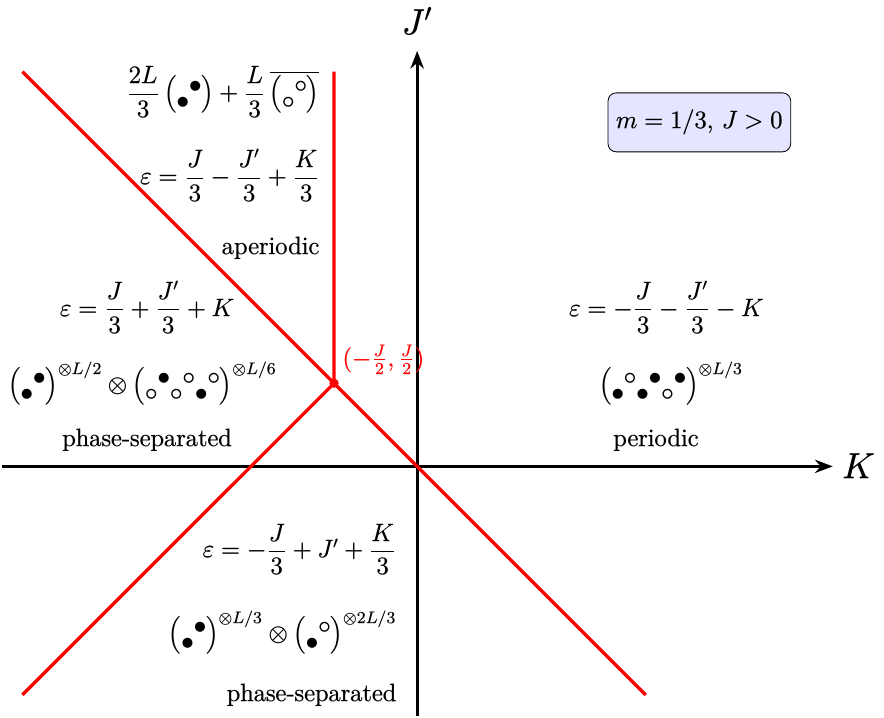}
    \end{subfigure}
\caption{Phase diagrams at $m=0$, $m=1/3$ for $J>0$ and $J<0$. Top row: $m=0$. Bottom row: $m=1/3$. Left column: $J<0$. Right column: $J>0$. In each region, only one representative state is depicted for clarity.
For an arbitrary magnetization value $m$, the phase diagram consists of phases with the same structure as those at ${m=0}$ and ${m=1/3}$, but with different number of particles. The boundaries between phases are accordingly shifted.}
\label{fig:PD}
\end{figure}
% ------------------------------------------------

The phase diagrams for ${m=0}$ and ${m=1/3}$ are shown in Fig.~\ref{fig:PD}. For any intermediate magnetization, ${0 < m < 1/3}$ or ${1/3 < m < 1}$, the phase diagram consists of phases with the same structure as those at ${m=0}$ and ${m=1/3}$, but with different number of particles. As $m$ increases, each phase in the ${m=0}$ diagram evolves smoothly into its counterpart at ${m=1/3}$, and subsequently, each of those phases continuously transforms into the ferromagnetic state at ${m=1}$. The boundaries between phases are accordingly shifted. These shifts give rise to first-order phase transitions when $m$ varies with other parameters fixed.

As an example, consider the ground state
$
(
    \raisebox{-0.5ex}{$\circ$}
    \raisebox{ 0.5ex}{$\circ$}
)^{\otimes L/4}
\otimes
(
    \raisebox{-0.5ex}{$\bullet$}
    \raisebox{ 0.5ex}{$\circ$}
    \raisebox{-0.5ex}{$\bullet$}
    \raisebox{ 0.5ex}{$\bullet$}
    \raisebox{-0.5ex}{$\circ$}
    \raisebox{ 0.5ex}{$\bullet$}
)^{\otimes L/4}
$
at ${m=0}$. For ${0<m<1/3}$, according to $m$-dependent LP solution, it evolves into
$
(
    \raisebox{-0.5ex}{$\circ$}
    \raisebox{ 0.5ex}{$\circ$}
)^{\otimes (1-3m) L/4}
\otimes
(
    \raisebox{-0.5ex}{$\bullet$}
    \raisebox{ 0.5ex}{$\circ$}
    \raisebox{-0.5ex}{$\bullet$}
    \raisebox{ 0.5ex}{$\bullet$}
    \raisebox{-0.5ex}{$\circ$}
    \raisebox{ 0.5ex}{$\bullet$}
)^{\otimes (1+m)L/4}
,$
which at ${m=1/3}$ becomes
$
(
    \raisebox{-0.5ex}{$\bullet$}
    \raisebox{ 0.5ex}{$\circ$}
    \raisebox{-0.5ex}{$\bullet$}
    \raisebox{ 0.5ex}{$\bullet$}
    \raisebox{-0.5ex}{$\circ$}
    \raisebox{ 0.5ex}{$\bullet$}
)^{\otimes L/3}
.$
As $m$ increases further, flipping any $\circ$ spin into $\bullet$ yields the same energy change, so the state becomes aperiodic in the range ${1/3 < m < 1}$.

\begin{figure}[t]
    \begin{subfigure}[b]{0.45\textwidth}
         \centering
         \includegraphics[width=\textwidth]{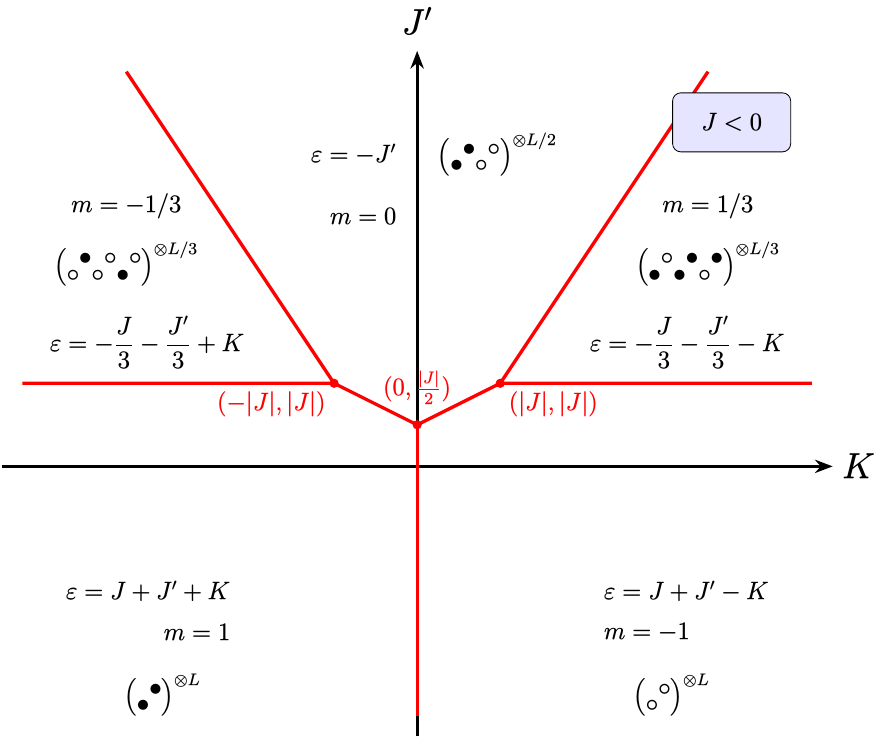}
    \end{subfigure}
    \hfill
    \begin{subfigure}[b]{0.45\textwidth}
         \centering
         \includegraphics[width=\textwidth]{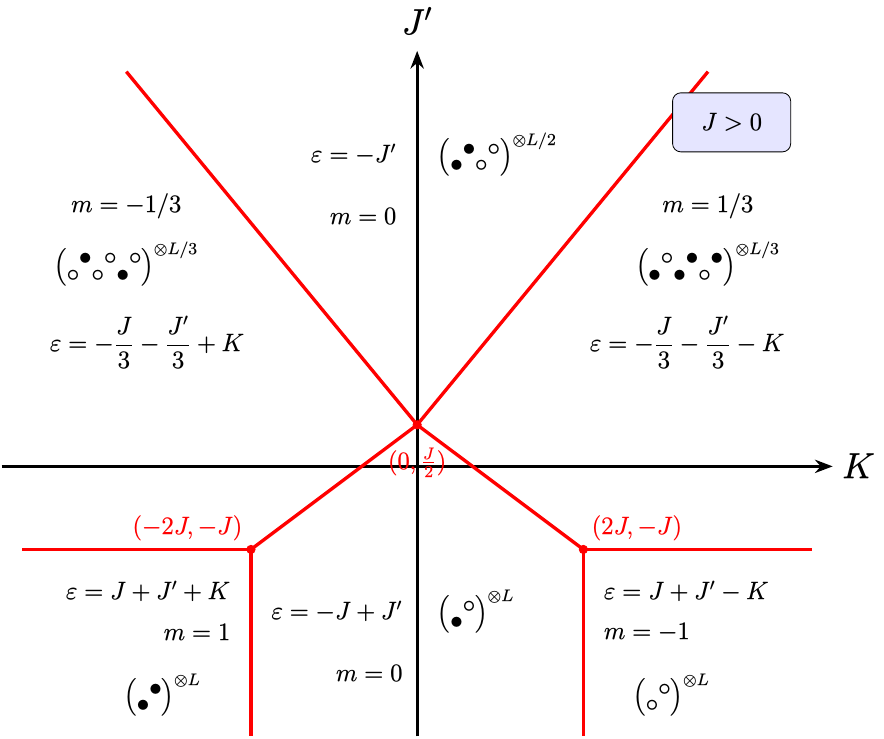}
    \end{subfigure}
\caption{Phase diagram for unrestricted magnetization. Only periodic ground states are realized, with $m$ taking one of the critical values, except for the phase boundaries.}
\label{fig:PD-m}
\end{figure}

Next, we analyze the problem from the perspective where the magnetization is treated as a free parameter. Within the LP formulation, this corresponds to including $m$ as the additional component of the vector $\bm{y}$. In this case, the ground state adopts only periodic configurations~\cite{Morita_1974_1}. 
Except for the phase boundaries, the average magnetization per site takes one of the values: $0$, $\pm 1/3$, or $\pm 1$; in agreement with the thermodynamical results obtained in Ref.~\cite{Jurcisinova_2014}.
The resulting phase diagram is shown in Fig.~\ref{fig:PD-m}.

Finally, we comment on the relation between the present results and the existing literature on the FKM. 
For small values of $K$ ${(|K| \ll |J|, |J'|)}$, the model~\eqref{eq:H} effectively describes the strong-coupling regime of the FKM on a triangular ladder.
Previous studies of the FKM on triangular lattices revealed a rich variety of competing ordering tendencies, such as stripe phases, phase separation, and frustration-driven complex patterns~\cite{2D_FKM_2007}. 
These phases are counterparts to the periodic, phase-separated, and ordered but aperiodic ground states identified in the present work. 
See also Refs.~\cite{2D_FKM_2011, Freericks_2002, Yadav_2010} for related studies of ordering phenomena in square and extended triangular FKM models.

% ------------------------------------------------
% ------------------------------------------------

% ------------------------------------------------
\section{Summary}
\label{sec:summary}

In this work we employed the LP approach to determine exactly the ground states of the Ising model at arbitrary fixed magnetization on a two-leg triangular ladder with three-spin interactions.
A general algorithmic method was formulated to identify all relevant constraints in the thermodynamic limit, including both equality and inequality relations.
The issue of ``inconstructible" vertices was resolved for the considered chain model, resulting in a complete enumeration of the ground states.

Our analysis revealed three distinct classes of ground-state configurations: periodic, phase-separated, and ordered but aperiodic. A periodic state is generated by a finite supercell; a phase-separated ground state consists of two regions, each forming a periodic structure with a distinct supercell; and an aperiodic state is composed of two different finite blocks arranged in arbitrary order, subject to certain local rules.

We identified the critical magnetization values ${m=0}$, ${m=\pm 1/3}$, and the trivial ${m=\pm 1}$, at which the feasible polytope changes its shape. These transitions correspond to qualitative changes in the phase diagram. We constructed the phase diagrams at ${m=0}$ and ${m=1/3}$, and demonstrated how the phase diagram evolves continuously as $m$ varies between the critical values.

When the magnetization is treated as a free parameter, the system selects only periodic ground states.
Phase diagram in case of unrestricted magnetization has also been constructed.

% ------------------------------------------------
\section*{Acknowledgments}
The author would like to thank George I. Japaridze, Alexander A. Nersesyan, and George Jackeli for valuable discussions.

% ------------------------------------------------
\appendix
\numberwithin{equation}{section}

% ------------------------------------------------
\section{Derivation of the parametrization}
\label{app:parametrization}

In this appendix, we present the derivation of the parametrization given in Eq.~\eqref{eq:parametrization},
which is most conveniently carried out in matrix form. 
To that end, Eq.~\eqref{eq:x-uv} can be written as
% ------------------------------------------------
\begin{align}
\label{eq:x-u_matrix}
    \bm{x} = A \bm{w} \,,
\end{align}
% ------------------------------------------------
where $A$ is a ${d_x{\times}d_w}$ matrix of rank $d_x$
and the $d_w{\times}1$ vector $\bm{w}$ was introduced in Eq.~\eqref{eq:w}.
The nine linearly independent constraints, whose selection is not unique, take the form
% ------------------------------------------------
\begin{align}
\label{eq:constraints_matrix}
    \bm{b} = B \bm{w} \,,
\end{align}
% ------------------------------------------------
where $\bm{b}$ is a ${d_b{\times}1}$ vector with ${d_b=9}$ and $B$ is a ${d_b{\times}d_w}$ matrix of rank $d_b$.
If one selects the following 9 linearly independent constraints: Eq.~\eqref{eq:m}, both normalizations in Eq.~\eqref{eq:sum_uv}, any three in Eq.~\eqref{eq:bot_R-top-L} and any three in Eq.~\eqref{eq:bot_L-top-R}, then $b_1 = m$, $b_2=b_3=1$ and $b_i=0$ for $4 \leqslant i \leqslant 9$.
Eqs.~\eqref{eq:x-u_matrix} and \eqref{eq:constraints_matrix} can be combined into a single equation by stacking vertically
% ------------------------------------------------
\begin{align}
\label{eq:x'-Sw}
    \bm{x}' = S \bm{w} \,,
\end{align}
% ------------------------------------------------
with
% ------------------------------------------------
\begin{align}
    \bm{x}' = 
    \begin{pmatrix}
        \bm{x} \\ \bm{b}
    \end{pmatrix}
,
\qquad
    S = 
    \begin{pmatrix}
        A \\ B
    \end{pmatrix}
,
\end{align}
% ------------------------------------------------
where $\bm{x}'$ is a ${(d_x+d_b){\times}1}$ vector and $S$ is a ${(d_x+d_b){\times}d_w}$ matrix of rank ${d_x+d_b}$.
Eq.~\eqref{eq:x'-Sw} can be inverted as
% ------------------------------------------------
\begin{align}
    \bm{w}
    = 
    S^{+}
    \bm{x}'
    +
    \mathcal{N}(S) \,
    \bm{y}
\,,
\end{align}
% ------------------------------------------------
where $S^{+}$ is the unique ${d_w{\times}(d_x+d_b)}$ Moore-Penrose pseudoinverse of $S$; the columns of the matrix $\mathcal{N}(S)$ form a basis for the null space of $S$; and $\bm{y}$ is an arbitrary ${d_y}{\times}1$ vector, where $d_y$ is the dimension of the null space of $S$. By the rank-nullity theorem, ${d_y = d_w - (d_x+d_b) = 4}$. Accordingly, $\mathcal{N}(S)$ is a ${d_w{\times}d_y}$ matrix.
The matrix $\mathcal{N}(S)$ is not uniquely defined, since the null-space basis of $S$ can be chosen arbitrarily. 
The choice leading to Eq.~\eqref{eq:parametrization} was described immediately before that equation.
While the derivation here focuses on Eq.~\eqref{eq:parametrization}, the same procedure can be applied systematically to other LP problems.

% ------------------------------------------------
\printbibliography

\end{document}